\documentclass[aps,twocolumn,showpacs,superscriptaddress,amsmath,amssymb,prl,nofootinbib]{revtex4-1}
\usepackage{epsf,amsmath,amssymb,amsfonts,verbatim,color,multirow,pifont}
\usepackage{graphicx}
\usepackage{dcolumn}
\usepackage{bm}
\usepackage{txfonts}
\usepackage{hyperref}
\usepackage{soul}

\begin{document}
\title{Carnot efficiency and zero-entropy-production rate do not guarantee reversibility of a process }
\author{Jae Sung Lee}
\affiliation{Quantum Universe Center, Korea Institute for Advanced Study, Seoul 02455, Korea}
\author{Sang Hoon Lee}
\affiliation{School of Physics, Korea Institute for Advanced Study, Seoul 02455, Korea}
\affiliation{Department of Liberal Arts, Gyeongnam National University of Science and Technology,  Jinju 52725, Korea}
\author{Jaegon Um}
\email{slung@postech.ac.kr}
\affiliation{BK21PLUS Physics Division, Pohang University of Science and Technology, Pohang 37673, Korea}
\author{Hyunggyu~Park}
\email{hgpark@kias.re.kr}
\affiliation{School of Physics, Korea Institute for Advanced Study, Seoul 02455, Korea}
\affiliation{Quantum Universe Center, Korea Institute for Advanced Study, Seoul 02455, Korea}

\newcommand{\revise}[1]{{\color{red}#1}}

\date{\today}

\begin{abstract}
Thermodynamic process at zero-entropy-production (EP) rate has been regarded as a reversible process. A process achieving the Carnot efficiency is also considered as a reversible process. Therefore, the condition, `Carnot efficiency at zero-EP rate' could be regarded as a strong equivalent condition for a reversible process. Here, however, we show that the detailed balance can be broken 
for a zero-EP rate process and even for a process achieving the Carnot efficiency at zero-EP rate in an example of a quantum-dot model. This clearly demonstrates that `Carnot efficiency at zero-EP rate' or just 'zero-EP rate' is not a sufficient condition for a reversible process.
\end{abstract}


\maketitle

\emph{Introduction} -- Reversible process is a process that its reversed one returns the system to the initial state without leaving any trace in environments. Therefore, for being a reversible process, every transition should be equilibrated by its reversed transition, which is called the detailed balance (DB). If the DB is satisfied, no current can flow. In addition, no entropy is produced in a DB-satisfied process, as the entropy production (EP) can be defined by the logarithmic ratio between forward and its time-reversal path probabilities~\cite{Udo_review}.

As a system should be always maintained in an equilibrium state during the process, it takes an infinitely long time to implement the process in an exactly reversible way. However, such an infinite-time process does not exist in the real world. Therefore, a reversible process is usually understood as a quasi-static-limit (very slowly varying) process for practical purpose. In this limit, all currents including the EP rate
should vanish. Therefore, in this context, the `zero-EP rate' limit has been usually and practically regarded as an equivalent condition for the reversible limit.

If we focus our discussion on heat engines working between two reservoirs at temperatures $T_1$ and $T_2$ ($T_1>T_2$), there is another conventional indicator for the reversibility: that is, how close the engine efficiency is to the Carnot efficiency $\eta_\textrm{C} = 1-T_2/T_1$. This ideal efficiency is attainable in a reversible process as in the well-known Carnot engine~\cite{book1}. This can be easily understood by the following relation for the efficiency $\eta$ and the total EP per engine cycle $\Delta S$~\cite{JSLee}:
\begin{equation}
\eta_\textrm{C} -\eta = \frac{T_2 \Delta S}{Q_1}, \label{eq:eta_relation}
\end{equation}
where $Q_1$ is the amount of heat absorbed from the hotter reservoir. From Eq.~\eqref{eq:eta_relation}, it is obvious that $\eta$ approaches $\eta_\textrm{C}$ in the $\Delta S \rightarrow 0$ (reversible) limit. On the other hand, for $\Delta S > 0$ (irreversible process), $\eta$ should be lowered as much as $T_\textrm{2} \Delta S /Q_1$.
For this reason, the limit achieving $\eta_\textrm{C}$ has been considered to be equivalent to  the reversible limit. 
However, note that the reversibility may not be required in achieving $\eta_\textrm{C}$ when $Q_1\rightarrow \infty$.

For a steady state engine, Eq.~\eqref{eq:eta_relation} can be written as
\begin{equation}
\eta_\textrm{C} -\eta=\frac{T_2 \dot{S}}{\dot{Q}_1}. \label{eq:steady_eta_relation}
\end{equation}
where $\dot{Q}_1$ and $\dot{S}$ are the steady state rates of  $Q_1$ and EP, respectively. In the reversible limit, $\dot{S}$ approaches zero and the Carnot efficiency is attained. Therefore, the $\dot{S}\rightarrow 0$ limit has been also regarded to guarantee the Carnot efficiency. Again, however, this may not be correct in some limits such as $\dot{Q_1} \rightarrow 0$ or $\infty$. Furthermore, the
$\dot{S}\rightarrow 0$ limit does not always guarantee the reversibility in the sense of the DB satisfiability (discussed later).
Nevertheless, it has been conventionally believed that these limits might be equivalent.

Recently, several studies pointed out that the conventional belief could be wrong by studying
explicit models violating the equivalence~\cite{Allahverdyan, JSLee, Polettini}.
Lee and Park~\cite{JSLee} showed that the efficiency of the Feynman-Smoluchowski ratchet~\cite{Smoluchowski, Feynman} can approach the Carnot bound with non-vanishing $\Delta S$ (DB violation) in a specific limit. In this ratchet model, the system should overcome a steep hill of the periodic energy barrier with height $U$ to extract work.
They found that $\Delta S \propto \ln U$ and $Q_1 \propto U$ when the system overcomes the energy barrier once (one engine cycle). Therefore, in the $U \rightarrow \infty$ limit, the Carnot efficiency is attainable from Eq.~\eqref{eq:eta_relation}.
However, as it takes $\sim e^U$ time for overcoming a barrier, the EP rate vanishes as $e^{-U} \ln U $ ($\dot{S}\rightarrow 0$). 
This limit is peculiar in that positive entropy is produced when overcoming an energy barrier, but all currents including the EP rate vanish due to the exponentially slow process. 
As vanishing currents are key features also in a reversible process, there was some confusion on whether this process is classified as an irreversible or a reversible one. To clear out the ambiguity in determining the reversibility, one should examine the DB satisfiability in the steady state; if the DB is broken, the process cannot be reversible~\cite{HKLee}. It turns out that the DB does not hold in this limit with
the zero-EP rate and the Carnot efficiency, which clearly shows that the conventional belief of the equivalence does not hold.

Polettini and Esposito~\cite{Polettini} showed that the Carnot efficiency can be attained at a divergent power output in a two-cycle model. In this limit, both $\dot{S}$ and $\dot{Q}_1$ diverge while the ratio $\dot{S} /\dot{Q}_1$ vanishes, thus
the efficiency approaches $\eta_\textrm{C}$ by Eq.~\eqref{eq:steady_eta_relation}. This confirms that the Carnot-efficiency limit does not guarantee the zero-EP rate and the reversibility.

\begin{figure}
\centering
\includegraphics[width=0.85\linewidth]{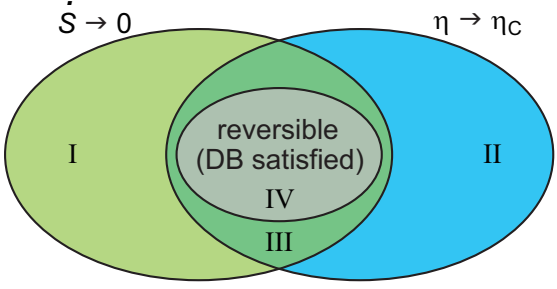}
\caption{(Color online) Venn diagram showing relations between the three limits. Here, the reversible limit refers to a limit process satisfying the DB condition.} \label{fig:venn}
\end{figure}

In this work, we study the relationship between $\dot{S} \rightarrow 0$, $\eta \rightarrow \eta_\textrm{C}$, and the reversibility
(DB satisfiability) in a systematic way.  The reversibility condition, of course, guarantees the two other limits of $\dot{S}\rightarrow 0$ and $\eta\rightarrow \eta_\textrm{C}$. Then, the most general logical Venn diagram for the three limits can be drawn as in Fig.~\ref{fig:venn},
which suggests four possible cases.
(i) Region I: The EP rate vanishes without the Carnot efficiency.
(ii) Region II: The Carnot efficiency is attained with non-vanishing EP rate. The model studied by Polettini and Esposito~\cite{Polettini} belongs to this case.
(iii) Region III: A process is irreversible even when both $\dot{S} \rightarrow 0$ and $\eta \rightarrow \eta_\textrm{C}$ are satisfied. Therefore, `Carnot efficiency at zero EP rate' does not guarantee a reversible process. The Feynman-Smoluchowski ratchet~\cite{JSLee} is one of such examples.
A similar behavior was also observed in other systems such as a quantum refrigerator~\cite{Allahverdyan}.
(iv) Region IV: All three limits are realized simultaneously, which corresponds to the conventional belief.

\begin{figure}
\centering
\includegraphics[width=0.9\linewidth]{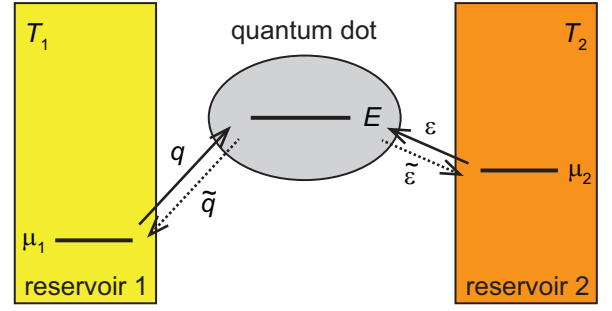}
\caption{(Color online) Schematic of the model. There are two reservoirs or leads $1$ and $2$ with temperatures $T_1$ and $T_2$ and chemical potentials $\mu_1$ and $\mu_2$, respectively. Electron can move from one reservoir to the other via the quantum dot which has a well-defined single energy level $E$. The transition rate from the reservoir $1$ ($2$) to the dot is $q$ ($\epsilon$) and the reversed rate is $\tilde{q}$ ($\tilde{\epsilon}$). } \label{fig:schematic}
\end{figure}

\emph{Model} -- To demonstrate our conclusion shown in Fig.~\ref{fig:venn}, we consider the following thermoelectric device~\cite{Esposito,Josefsson,LUP} as illustrated in Fig.~\ref{fig:schematic}. This device consists of a quantum-dot which is in contact with two leads or reservoirs with different temperatures $T_1$ and $T_2$ and different chemical potentials $\mu_1$ and $\mu_2$, respectively. Electrons can move from one reservoir to another via the quantum dot where only a single electron can be occupied at a state with a sharply defined energy $E$. Thus, there are two states of the dot: occupied and unoccupied states whose energies are $E$ and $0$, respectively.

In this study, we consider the case $T_1>T_2$ and $\mu_1 < \mu_2< E$. The transition rate of an electron from the lead $1$ ($2$) to the dot is $q$ ($\epsilon$) and the corresponding reverse rate is $\tilde{q}$ ($\tilde{\epsilon}$). Then, this system can be described by the following master equation~\cite{Esposito,LUP,Bonet,Bagrets,Harbola}:
\begin{align}
&\dot{P}_\textrm{oc} = (q+\epsilon) P_\textrm{un} - (\tilde{q}+\tilde{\epsilon}) P_\textrm{oc} \nonumber \\
&\dot{P}_\textrm{un} =  (\tilde{q}+\tilde{\epsilon}) P_\textrm{oc} - (q+\epsilon) P_\textrm{un} , \label{eq:master_eq}
\end{align}
where $P_\textrm{oc}$ and $P_\textrm{un} $ are probabilities of occupied and unoccupied states of the quantum dot, respectively. Here, we assume the local detailed balance conditions for the transition rates such that
\begin{align}
\frac{q}{\tilde{q}} &= e^{-(E-\mu_1)/T_1} \equiv x, \label{eq:x} \\
\frac{\epsilon}{\tilde{\epsilon}} &= e^{-(E-\mu_2)/T_2} \equiv y. \label{eq:y}
\end{align}
For simplicity, we set the time constants for the transition rates as $q+\tilde{q}=\epsilon+\tilde{\epsilon}=1$. Then, the steady-state solution of the master equation~\eqref{eq:master_eq} is given by
\begin{align}
P_\textrm{oc}^\textrm{ss} = \frac{1}{2} (q+\epsilon) = 1-P_\textrm{un}^\textrm{ss}, \label{eq:ss_sln}
\end{align}
where $P_\textrm{oc}^\textrm{ss}$ and $P_\textrm{un}^\textrm{ss}$ are steady-state probabilities of occupied and unoccupied states, respectively.
Then, the steady-state current of electrons becomes
\begin{align}
J^\textrm{ss}= q P_\textrm{un}^\textrm{ss} - \tilde{q}P_\textrm{oc}^\textrm{ss}  =  \frac{1}{2} \left(q-\epsilon\right)=\frac{x-y}{2(1+x)(1+y)}.  \label{eq:ss_current}
\end{align}
We note that the DB condition for the probabilistic current balance
between the quantum dot and each lead reads as
\begin{align}
\frac{q P_\textrm{un}^\textrm{ss}}{\tilde{q}P_\textrm{oc}^\textrm{ss}}  =
\frac{\epsilon P_\textrm{un}^\textrm{ss}}{\tilde{\epsilon}P_\textrm{oc}^\textrm{ss}} = 1
\quad  \Longleftrightarrow \quad x=y \quad \textrm{(DB condition)}~,  \label{eq:DB}
\end{align}
with which we get $J^\textrm{ss}=0$ trivially.

As the energy and matter are strongly coupled in this model, the steady-state heat currents are given as follows:
\begin{align}
\dot{Q}_1 = J^\textrm{ss}(E-\mu_1)=-J^\textrm{ss} T_1 \ln x \label{eq:ss_heat1}, \\
\dot{Q}_2 = J^\textrm{ss}(E-\mu_2)=-J^\textrm{ss} T_2 \ln y \label{eq:ss_heat2},
\end{align}
where $\dot{Q}_1$ ($\dot{Q}_2$) is the heat current from the lead $1$ ($2$) to the dot, respectively. Then, the work rate is the difference between the two heat currents:
\begin{align}
\dot{W} = \dot{Q}_1 -\dot{Q}_2 = J^\textrm{ss} (T_2 \ln y - T_1 \ln x) .  \label{eq:ss_work}
\end{align}
By definition, positive $\dot{W}$ means useful work extraction as an engine.
Using Eqs.~\eqref{eq:ss_heat1} and \eqref{eq:ss_heat2}, we can calculate the EP rate $\dot{S}$ and the efficiency $\eta$ as
\begin{align}
\dot{S} &= \frac{\dot{Q}_2}{T_2} -\frac{\dot{Q}_1}{T_1} = J^\textrm{ss} (\ln x -\ln y ), \label{eq:EPrate} \\
\eta &= 1 -\frac{\dot{Q}_2}{\dot{Q}_1} = 1-\frac{T_2 \ln y}{T_1 \ln x} \label{eq:eta}.
\end{align}

\begin{figure}
\centering
\includegraphics[width=0.7\linewidth]{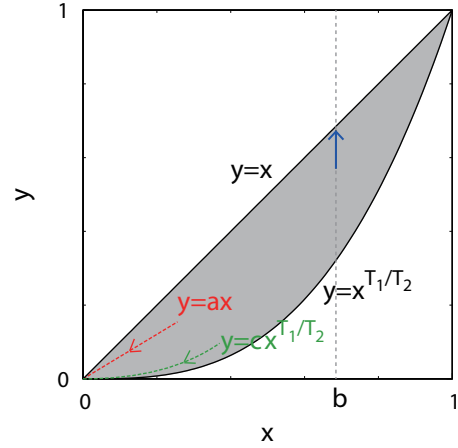}
\caption{(Color online) $x-y$ diagram. The shaded area is the region satisfying Eq.~\eqref{eq:xy_condition}. There are various limits to reach the equilibrium line ($y=x$) and the origin ($x=y=0$).} \label{fig:xy_condition}
\end{figure}

\emph{Various limit processes} -- There are two conditions for $x$ and $y$: (i)
From the thermodynamic second law, $\dot{S}\geq 0$ and (ii) for being a useful engine, $\dot{W} \geq 0$. These two conditions are summarized as
\begin{align}
x^{T_1/T_2} \leq y \leq x \label{eq:xy_condition}
\end{align}
and the corresponding region is shaded in Fig.~\ref{fig:xy_condition}. The lower bound ($y=x^{T_1/T_2}$) corresponds to
$\dot{W} =0$ ($\mu_1 = \mu_2$), $\eta = 0$,  $J^{\textrm{ss}} > 0$ and $\dot{S} > 0$, and the upper bound ($y=x$) corresponds to
equilibrium with $J^{\textrm{ss}} = 0$ and thus $\dot{S} = \dot{W} = \dot{Q_1} =\dot{Q_2}=0$.

We first consider a simple limit to reach the equilibrium  line ($y=x$) for fixed nonzero $x$. This can be realized by varying $\mu_2$ close to $\mu_1+ \eta_\textrm{C} (E-\mu_1)$ with fixed $E$ and $\mu_1$. This is the {\em reversible} limit, where $\eta \rightarrow \eta_\textrm{C}$
 and $\dot{S} \rightarrow 0$ with the DB condition Eq.~\eqref{eq:DB} satisfied. In fact, any linear or nonlinear approach to the equilibrium line except for the origin ($x=y=0$) turns out to be  the reversible limit (Region IV).

However, there can be various limits possible to approach the origin, where the energy gap $E-\mu_1$ ($E-\mu_2$) is much higher than
the thermal energy $T_1$ ($T_2$), respectively (see Eqs.~\eqref{eq:x} and \eqref{eq:y}). For example, if one approaches the origin along
the equilibrium line, the process is maintained as a reversible process with the Carnot efficiency (Region IV). The other simple limit is obtained by taking the lower-bound line ($y=x^{T_1/T_2}$) in Fig.~\ref{fig:xy_condition}.
Along this line, the efficiency is always zero ($\eta=0$)
and the DB is always broken ($x\neq y$).
The EP rate $\dot{S}$ vanishes as one approaches the origin because $J^\textrm{ss}$ vanishes faster, even though $\ln (x/y)$ diverges in Eq.~\eqref{eq:EPrate}. Note that $\ln (x/y)=\dot{S}/J^\textrm{ss}\equiv \Delta S$ represents the {\em average} EP per one electron transfer
(per one engine cycle), which is nonzero and in fact diverges in this limit. This is in sharp contrast to the former case (equilibrium line), where $\Delta S=\ln (x/y)=0$ due to the DB.
Thus, the latter case should be regarded as irreversible and belongs to Region I.

\begin{figure}
\centering
\includegraphics[width=0.99\linewidth]{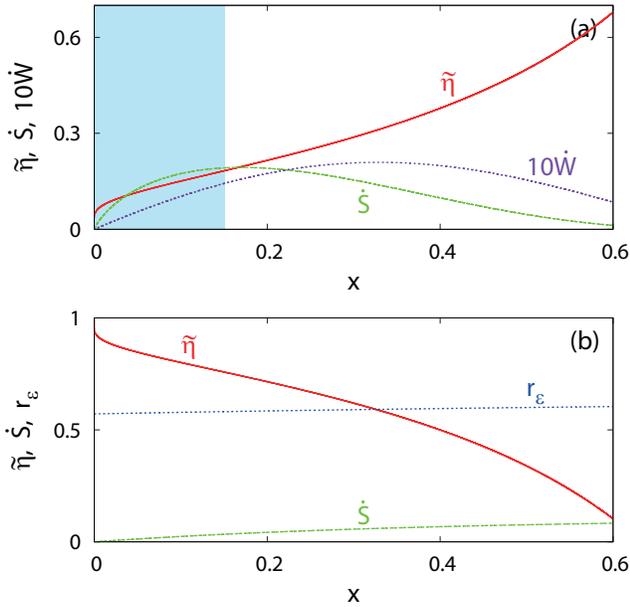}
\caption{(Color online) Normalized efficiency $\tilde{\eta}$, the EP rate $\dot{S}$, and the DB ratio $r_\epsilon$ as a function of $x$
(a) in the linear limit ($y=ax$) with $a=0.4$. As $x \rightarrow 0$, we find $\tilde\eta \rightarrow 1$, $\dot{S} \rightarrow 0$, and $r_\epsilon \rightarrow 4/7$. (b) Along the path $y=cx^{T_/T_2}$ with $c=XX$, we find $\tilde\eta \rightarrow 0$, $\dot{S} \rightarrow 0$, and $r_\epsilon \rightarrow 0$ as $x \rightarrow 0$. The blue shaded region indicates an anomalous behavior of the efficiency versus the EP rate. } \label{fig:limit}
\end{figure}

These two boundary limits are not useful, though, because the extracted power is always zero along boundaries ($\dot{W}=0$). We consider
other limits approaching the origin in between two boundaries. The simplest one is a {\em linear} limit along the $y=ax$ line with
$0<a<1$ as illustrated in Fig.~\ref{fig:xy_condition}. This can be achieved by tuning both energy gaps appropriately with fixed temperatures.
In this limit, one can see easily from Eqs.~\eqref{eq:ss_current}, \eqref{eq:ss_work}, \eqref{eq:EPrate}, and \eqref{eq:eta} that
$J^{\textrm{ss}} \rightarrow 0$, $\dot{W} \rightarrow 0$, $\dot{S} \rightarrow 0$, and $\eta \rightarrow \eta_\textrm{C}$.
With the zero-EP rate and the Carnot efficiency, this limit might be considered as a reversible limit.
Surprisingly, however, the DB conditions in Eq.~\eqref{eq:DB} is violated as
\begin{align}
 r_q\equiv \frac{q P_\textrm{un}^\textrm{ss} }{\tilde{q} P_\textrm{oc}^\textrm{ss} }= \frac{x(x+y+2)}{2xy+x+y}
\xrightarrow{x\rightarrow 0} \frac{2}{1+a} \neq 1, \nonumber \\
 r_\epsilon\equiv\frac{ \epsilon P_\textrm{un}^\textrm{ss}}{\tilde{\epsilon} P_\textrm{oc}^\textrm{ss} } = \frac{y(x+y+2)}{2xy+x+y}
 \xrightarrow{x\rightarrow 0}  \frac{2a}{1+a} \neq 1. \label{eq:DBcondition}
\end{align}
In Fig.~\ref{fig:limit}(a), $\dot{S}$, the normalized efficiency $\tilde{\eta}=\eta/\eta_\textrm{C}$,
and the probability current ratio $r_\epsilon$ are presented as a function of $x$ when $a=0.4$, $T_1 =1$, and $T_2 = 1/3$.
This clearly shows an example with both the Carnot efficiency and the zero-EP rate, but with the DB violated.
This limit belongs to Region III and should be regarded as irreversible.
The EP per cycle is finite such that $\Delta S = -\ln a >0$.

A more practical limit can be obtained by taking the $y=cx^{T_1/T_2}$ line with $c>1$.
This can be easily realized in experiments by increasing $E$ through controlling the gate voltage connected to the quantum dot with fixed
$\mu_1$ and $\mu_2$~\cite{Josefsson}. The results are similar to the simple lower-bound line case ($c=1$) such that
$J^{\textrm{ss}} \rightarrow 0$, $\dot{W} \rightarrow 0$, $\dot{S} \rightarrow 0$, and $\eta \rightarrow 0$. The current ratios
in this case $r_q\rightarrow 2$ and $r_\epsilon \rightarrow 0$ (broken DB), thus this limit belongs to Region I with diverging $\Delta S$.
Figure \ref{fig:limit} (b) shows the plot of various quantities when $c=2$, $T_1=1$, and $T_2=1/3$.
We note an interesting anomalous behavior such that `the larger irreversibility, the higher efficiency (or higher power)'. Usually, the efficiency or power decreases as the EP rate increases. However, in the blue shaded region of Fig.~\ref{fig:limit}(a), we can see the opposite behavior,
which was also reported previously in the Feynman-Smoluchowski ratchet~\cite{JSLee}.
For a general path limit with $x \sim y^\alpha$, we find that $J^{\textrm{ss}} \rightarrow 0$, $\dot{W} \rightarrow 0$, $\dot{S} \rightarrow 0$, and $\eta \rightarrow 1-(1-\eta_\textrm{C})\alpha$ with $r_q\rightarrow 2$, $r_\epsilon \rightarrow 0$, and
$\Delta S\rightarrow \infty$, which belongs to Region I.

The mechanism of this abnormal behavior is as follows. The EP rate can be factorized into two terms: $J^\textrm{ss}$ and $\Delta S$, that is,
\begin{align}
\dot{S} = J^\textrm{ss} \Delta S  \label{eq:two_parts},
\end{align}
where $\Delta S$ is the EP per engine cycle. With the DB satisfied, the EP is always zero by definition ($\Delta S=0$) and thus $\dot{S}=0$, which is the usual reversible limit.
However, one can reach the zero-EP rate with the DB violated ($\Delta S \neq 0$), when the engine is operated so slowly that $J^\textrm{ss}$
vanishes in some limits. This case generally belongs to Region I.

In some special cases, the efficiency may also reach the Carnot
efficiency, when the EP rate vanishes significantly faster than the heat absorption rate, $\dot{S}\ll \dot{Q}_1$ (see Eq.~\eqref{eq:steady_eta_relation}). This case belongs to Region III and was found in the linear limit of the quantum dot model
with non-zero finite $\Delta S$ and diverging $\Delta Q_1$ (heat absorbed per cycle) in Eq.~\eqref{eq:ss_heat1}.
This mechanism is essentially the same as what was found in the previous work~\cite{JSLee}, where
$\Delta S$ is also diverging but weaker (logarithmic divergence) than $\Delta Q_1$ (linear divergence) with increasing the energy barrier height.

\emph{Conclusion} -- In summary, we demonstrate that (i) the zero-EP limit does not guarantee the ideal efficiency nor the reversibility and (ii) it is possible to approach the Carnot efficiency at the zero-EP rate in an irreversible process. Using a simple quantum-dot model, we find that such a limit can be achieved by properly increasing the energy of the quantum dot or the chemical potentials. The $\dot{S} \rightarrow 0$ and $\eta \rightarrow \eta_\textrm{C}$ limit is also consistent with the recently proven power-efficiency trade-off relations~\cite{Shiraish, Pietzonka,Dechant}, in that the power vanishes with the Carnot efficiency.

Finally, we add a comment about an experiment on the quantum-dot model. The quantum dot model was experimentally implemented by using the setup studied by Josefsson \emph{et al}.~\cite{Josefsson}. In this experiment, the charging energy of the quantum dot is $4.9$ meV at $T_1=2$ K (corresponding thermal energy is $0.17$ meV). Thus, $x$ can be reduced to $ \sim e^{-4.9/0.17} \approx 3.0\times 10^{-13}$. If we take $T_2 = 1$ K and $a=1/e$, the normalized efficiency $\tilde\eta$ becomes $0.97$ with vanishing EP rate and broken DB. Therefore, the limits considered in this work could be  accessible in the real experiments.

\begin{acknowledgments}
This research was supported by the NRF Grant No.~2011-35B-C00014 (JSL), No.~2018R1C1B5083863 (SHL),
No.~2017R1D1A1B03030872 (JU), and No.~2017R1D1A1B06035497 (HP).
\end{acknowledgments}

\vfil\eject

\end{document}